\documentclass[11pt,twoside]{article}
\usepackage{version}
\usepackage{amsmath}
\usepackage{ihepconf}
\newcommand{\beq}{\begin{equation}}
\newcommand{\eeq}{\end{equation}}
\newcommand{\Th}{{\Theta}}
\newcommand{\om}{{\omega}}
\def\B{S}
\def\M{{\cal M}}

\def\P{{\cal P}}
\def\H{{\cal H}}

\def\i{\gamma}

\def\ut#1{\rlap{\lower1ex\hbox{$\sim$}}{#1}}

\def\l{\ell_{Pl}}
\def\C{{\rm C}}
\def\U{{\rm U}}
\def\SU{{\rm SU}}

\def\SL{{\rm SL}}
\def\E{{}^\i\!\Sigma}
\def\A{{}^\i\!\!A}
\def\ba{\begin{eqnarray}}
\def\ea{\end{eqnarray}}
\def\be{\begin{equation}}
\def\ee{\end{equation}}
\begin{document}
\title{{BLACK HOLE STATISTICAL PHYSICS: ENTROPY}}
\author{{\bf Vladimir O. Soloviev}\\
{\it Institute for High Energy Physics, Protvino, Russia}}
\date{}
\maketitle
\vspace*{-5mm}

\begin{abstract}
We discuss the most interesting approaches to derivation of the Bekenstein-Hawking entropy formula from a
statistical theory.
\end{abstract}

\section{Introduction}

Black hole is one of new physical phenomena predicted by General Relativity (GR). Though the existence of black
holes is still hypothetical (see, for example,~\cite{Karczmarek}), they send a challenge to our imagination and
rise a lot of interesting questions. The space-time having a black hole in it, first, has a singularity, and
second, has a horizon preventing an external observer from seing it. The singularity in GR is radically different
from field theory singularities because it is a property not of some field  but of the space-time itself. The
topology of space-time is changed when it acquires a black hole.

The most convincing viewpoint is that arising of singularities demonstrate the incompleteness of classical gravitational
theory
and the need to replace it by the quantum theory. So, if one believes in GR as a right theory of gravitation then
he can assert that the singularity problem will be solved only after discovering of the quantum gravity theory.
But\ldots might quantum gravity be the End of Physics? If the problem of black hole singularity would be solved
then probably the problem of cosmological singularity will be solved also. And so, we will have an answer to the
question how, when and from what kind of material our Universe has been created. Then all physicists together should
migrate, maybe, to biology.

But sensible people suggest that the End of Physics is still far from us, and so, we are still far from quantum
gravity. Then what we can expect now? Maybe, one should try to guess at least one formula of the future mysterious
theory, like  Johann Jakob Balmer~\cite{Balmer}, Swiss teacher of mathematics at a secondary school for girls in
Basel, has done in 1885. Or, if one believes that such a formula has been already discovered by
Bekenstein~\cite{Bekenstein:72,Bekenstein:73,Bekenstein:74}, one should hope to derive it from some intermediate
between classical and quantum theories model, like Niels Bohr has done in 1913 with his model of hydrogen atom.

Thorough study of black hole solutions and their properties such as test particle behavior has led to formulation
of ``black hole thermodynamics''~\cite{BCH:73}, i.e.  four laws which are very similar to four laws of the
thermodynamics:

{\bf 0. Surface gravity $\kappa$ is constant over the horizon.}

{\bf 1. Energy is conserved.}
$$\delta M=\frac{\kappa\delta A}{8\pi}+\Omega\delta J+\Phi\delta Q$$

{\bf 2. Matter entropy plus sum of areas of all horizons can not decrease.}
$$
\frac{d}{dt}\left(S_{matter}+S_{BH} \right)\ge 0 $$

{\bf 3. It is impossible to reduce surface gravity to zero by a finite sequence of operations.}

This discovery was motivated by an observation that black holes revealed ir\-re\-ver\-si\-bi\-li\-ty: a particle
falling into the black hole was unable to be back. This particle enlarges the black hole mass and also the
gravitational radius and the horizon area. It is impossible to decrease these quantities in classical
physics~\cite{Hawking:71,Hawking:72}.

The entropy is an example of non-decreasing variable in the standard thermodynamics.
Be\-ken\-stein~\cite{Bekenstein:72,Bekenstein:73,Bekenstein:74} attributed entropy to black holes  by suggestion
that it is proportional to the horizon area. The constant of proportionality was determined by Be\-ken\-stein and
Hawking in articles~\cite{Bekenstein:74,Hawking:74,Hawking:75}. So, the following formula has appeared
\begin{equation}
S_{BH}=A/4G=A/4\ell^2_{Pl} \ ,\label{entropy_BH}
\end{equation}
where $S_{BH}$ is the Bekenstein-Hawking entropy and $\ell_{Pl}=\sqrt{\hbar G/c^3}$ is the Planck length.

Another characteristic variable of thermodynamics is the temperature. Black holes also have analogous variable.
This is the surface gravity $\kappa$ which (in case of a Killing horizon) is defined by the following equation
\begin{equation}
\chi^c\nabla_c\chi_b=\kappa\chi_b,
\end{equation}
where  $\chi^a$ is a null Killing vector tangential to the horizon. Practically this temperature is rather low for
black holes which could be formed as a result of  a star collapse. If  $M_{\odot}$ is  mass of the Sun then in
standard Kelvin units
\begin{equation}
T_{BH}=\kappa/2\pi=\frac{\hbar}{8\pi k_B }M^{-1}\approx
10^{-7}M_{\odot}/M.
\end{equation}

Nevertheless Hawking suggested that taking a black hole together with radiation of the same temperature and
putting them into a thermostat should provide a system in thermodynamical equilibrium. Therefore the black hole
absorbing radiation have also to emit. Under some conditions Hawking has predicted~\cite{Hawking:75} that a process of
emitting particles
takes place for any black hole. Here the gravitational field of black hole is treated as a classical background whereas
the field of
 emitted particles is quantized.

The Hawking radiation becomes relatively stronger if the black hole becomes smaller. But the back-reaction of
emitted particles on the gravitational field was neglected. Hawking had enough courage to say that
a black hole could even explode or evaporate. This suggestion has led to a new puzzle: if a black hole
disappears then what can be said about the information ``eaten'' by this hole? What if unitarity or even quantum
mechanics are violated by this process? This is an excited and open question (sometimes called ``black hole
information puzzle''), but we will not discuss it here (see~\cite{Hawking:2004}).

The subject of this talk is a review of statistical derivations of the black hole entropy formula. History of
physics shows that all thermodynamical laws later have been derived from a more fundamental theory --- statistical
mechanics. It was shown that the concept of temperature could be explained as the average kinetic energy of a
micro-particle, the concept of entropy --- as logarithm of the number of states corresponding to the same
macroscopic thermodynamical macro-state. Looking at the black hole from this viewpoint we are to relate its
entropy with the number of states of some quantum theory. If we consider Bekenstein-Hawking formula in some sense
analogous to the Balmer formula for hydrogen atom then it is required to construct a phenomenological model of
quantized black hole which would lead to Eq.~(\ref{entropy_BH}).

It is natural that there are different approaches to this problem. Which one  is true or maybe all of them are
 misleading, nobody knows now. Below we demonstrate in short the most popular methods and their achievements. Personal
 sympathies of the author can be seen from  proportions of text devoted to one or another approach.

\section{Conformal field theory}
\addcontentsline{toc}{section}{Conformal field theory}

\subsection{Thermodynamics of almost extremal black holes}
\addcontentsline{toc}{subsection}{Black holes which look two-dimensional}

There is an important analogy described by  Fursaev~\cite{Fursaev}: thermodynamical properties of almost extremal
black hole correspond to those of a gas of massless scalar particles in flat 1+1-space. We start with explanation
of this idea below.

A charged black hole is  described by Reissner-Nordstr\"om solutions
\begin{equation}\label{1.1}
ds^2=-Bdt^2+\frac{dr ^2}{B}+r^2d\Omega^2,
\end{equation}
where $d\Omega^2$ is the metric on a unit sphere,
\begin{equation}\label{1.2}
B=\frac{1}{r^2}(r-r_-)(r-r_+),~~r_{\pm}=m\pm\sqrt{m^2-q^2},
\end{equation}
$q=Q\sqrt{G}$ is related to the electric charge $Q$ of the black hole, and $m=MG$ to its mass $M$. The radius of
the horizon is $r_+$. The Hawking temperature  of this black hole is
\begin{equation}\label{1.3}
T_H=\frac{1}{2\pi r_+}\sqrt{m^2-q^2},
\end{equation}
and the Bekenstein-Hawking entropy is $S_{BH}=\pi r_+^2/G$.

This solution has an interesting property: the Hawking temperature vanishes  when $m\rightarrow q$.  A solution
with $m=q$ is called an extremal black hole. There are no physical processes allowing  to convert a black hole
with $m>q$ into an extremal one. Nevertheless the extremal black holes acquired a lot of theoretical attention.

Let us consider now black holes which are ``almost extremal''
\begin{equation}\label{1.4}
m=q+E,~~~E \ll q.
\end{equation}
Thermodynamical relations for these objects are
\begin{equation}\label{1.6}
E=m-q=\lambda^2 T_H^2,~~ S=S_{BH}-\frac{\pi}{G} q^2=\frac{2\lambda^2}{G} T_H,
\end{equation}
where $\lambda = (2\pi^2 q^3)^{1/2}$.

Now  consider a gas of massless non-interacting scalar fields $\phi_k$, $k=1,\ldots,c$ on an interval $[0,b]$. The
dynamical equations and boundary conditions are as follows
\begin{equation}\label{1.8}
(\partial_t^2-\partial_x^2)\phi_k(t,x)=0,~~~ \phi_k(t,0)=\phi_k(t,b)=0.
\end{equation}
Let this system be in a state of thermal equilibrium at some temperature $T$. Then it is a one-dimensional analog
of an ideal gas of photons in a box.  The free energy of this gas
\begin{equation}\label{1.9}
F(T,L)=cT\sum_n\ln\left(1-e^{-\omega_n/T}\right),
\end{equation}
in the thermodynamical limit, $Tb\gg 1$,  can be  estimated as
\begin{equation}\label{1.10}
F(T,b)\simeq -\frac{\pi c}{6}bT^2,
\end{equation}
so, the energy $E(T,b)$ and the entropy $S(T,b)$ of this system are the following
\begin{equation}\label{1.7}
E(T,b) \simeq \frac{\pi c}{6}bT^2,~~ S(T,b)\simeq \frac{\pi c}{3}bT.
\end{equation}
By comparing (\ref{1.7}) with (\ref{1.6}) it is easy to see  that thermodynamical properties of an almost extremal
black hole are identical to properties of an ideal massless gas in a flat two-dimensional space-time.

\subsection{BTZ black holes}
\addcontentsline{toc}{subsection}{BTZ black holes}
The first statistical derivation of the black hole entropy formula based on
the conformal field theory (CFT) was suggested by Strominger~\cite{Stro:97}. We will follow him below in this subsection.

Consider  3-dimensional  GR with a negative cosmological constant $\Lambda=-l^{-2}$:
\begin{equation}\label{2.0}
I=\frac{1}{16\pi G}\int\sqrt{-g}d^3x\left(R+\frac{2}{\ell^2}\right).
\end{equation}
One of solutions for the corresponding gravitational equations is anti-de Sitter $AdS_3$ space:
\begin{equation}\label{2.1}
ds^2=-\left(1+\frac{r^2}{\ell^2}\right)dt^2+ \left(1+\frac{r^2}{\ell^2}\right)^{-1}dr^2+r^2d\varphi^2,
\end{equation}
where $\varphi$ has period $2\pi$. There is another exact solution discovered by  Ba\~nados, Teitelboim and
Zanelli~\cite{BTZ} and later called the BTZ black hole (here we consider only  non-rotating case):
\begin{equation}\label{2.2}
ds^2=-\frac{r^2-r_+^2}{\ell^2}dt^2+ \left(\frac{r^2-r_+^2}{\ell^2}\right)^{-1}dr^2+r^2d\varphi^2.
\end{equation}
Brown and Henneaux~\cite{BrHe} studied  a group of the asymptotic transformations, i.e. the transformations
preserving the property of a metric to be ``asymptotically anti-de Sitter''. The BTZ geometry is asymptotically
$AdS_3$ and fulfils the following boundary conditions first given in paper~\cite{BrHe}: \beq\label{2.222}
\begin{split}
g_{tt}&=
\frac{-r^2}{\ell^2} + {\cal O} (1),\\ g_{t\phi}&=
{\cal O}(1),\\
g_{tr}&= {\cal O} (\frac{1}{r^3}),\\
g_{rr}&= \frac{\ell^2}{r^2} + {\cal O} (\frac{1}{r^4}),\\ g_{r\phi}&={\cal
O} (\frac{1}{r^3}),\\ g_{\phi\phi}&= r^2 + {\cal O} (1)\ .\\
\end{split}
\eeq
Strominger has shown~\cite{Stro:97} that at
large $r$ the diffeomorphism vector fields  $\delta x^\mu=\zeta^\mu(x)$  preserving this asymptotic structure
(\ref{2.222}) are as follows:
\begin{equation}\label{2.5}
\zeta^t=\ell(T^+ +T^-)+\frac{\ell^3}{2r^2}(\partial_+^2T^+ +\partial_-^2T^-)+O(r^{-4}),
\end{equation}
\begin{equation}\label{2.6}
\zeta^\varphi=(T^+ -T^-)-\frac{\ell^2}{2r^2}(\partial_+^2T^- -\partial_-^2T^-)+O(r^{-4}),
\end{equation}
\begin{equation}\label{2.7}
\zeta^r=-\frac 12 r(\partial_+T^+ +\partial_+T^+)+O(r^{-1}),
\end{equation}
where $\partial_{\pm}=\ell\partial_t \pm \partial_\varphi$, $T^{+}$ and  $T^{-}$ are functions of one
variable, $t/\ell+\varphi$ and $t/\ell-\varphi$
 correspondingly.
The algebra of these vector fields, where the commutator, or a Lie bracket, $[\zeta_1,\zeta_2]$ of two vector
fields, $\zeta_1^\mu$ and $\zeta_2^\mu$, is a vector field with components $\zeta_3^\mu=\zeta_1^\nu\partial_\nu
\zeta_2^\mu- \zeta_2^\nu\partial_\nu \zeta_1^\mu$, preserves the asymptotics. The commutator of vector fields
$\zeta_1$ and $\zeta_2$ with the above asymptotic behavior is a vector field $\zeta_3$ which has the same
asymptotic behavior, and $T_3^{\pm}=T_1^{\pm}\partial_{\pm}T_2^{\pm}- T_2^{\pm}\partial_{\pm}T_1^{\pm}$. So,
generators of these diffeomorphisms should  form a closed algebra. Due to the periodicity over variable $\varphi$
and the specific dependence of functions $T^\pm$ on $t$ these functions also are periodic over time coordinate
with period~$2\pi\ell$.

Let $\zeta_n$, $\bar{\zeta}_n$ be defined by relations (\ref{2.5})--(\ref{2.7}) with
\begin{equation}\label{2.13}
T_n^{+}=\frac i2 e^{i n(t/\ell+\varphi)},~~ T_n^{-}=0,~~ \bar{T}_n^{+}=0,~~ \bar{T}_n^{-}=\frac i2 e^{i
n(t/\ell-\varphi)}.
\end{equation}
The algebra of these vector fields is as follows
\begin{equation}\label{2.8}
[\zeta_n,\zeta_m]=(n-m)\zeta_{n+m},~~ [\bar{\zeta}_n,\bar{\zeta}_m]=(n-m)\bar{\zeta}_{n+m},~~
[\zeta_n,\bar{\zeta}_m]=0.
\end{equation}
 and therefore the asymptotic symmetry group (\ref{2.5})--(\ref{2.7}) generated by two copies of the Virasoro algebra is
 the conformal group. So a consistent quantum gravity theory on $AdS_3$ should be a conformal field theory. It is well
 known that due to conformal anomaly  the Virasoro algebra has  a central extension. Surprisingly Brown and Henneaux
have derived this result even classically from surface terms appearing in Poisson brackets of the gravity
Hamiltonians.

In the Hamiltonian formalism of GR any infinitesimal diffeomorphism
 $\delta x^\mu=\zeta^\mu(x)$ is generated by a Hamiltonian of the following form
\begin{equation}\label{2.11}
H[\zeta]=\int_{\Sigma_t}\zeta^\mu {\cal H}_\mu d\Sigma+J[\zeta].
\end{equation}
This expression is a sum of a volume integral of a linear combination of constraints  ${\cal H}_\mu=0$ with the
diffeomorphism parameters as their
coefficients, and  a surface integral depending on the same parameters and on
canonical variables, $g_{ij}$, and $\pi_{ij}$.
 The surface term $J[\zeta]$ is introduced to ensure the canonical form for variations of
$H[\zeta]$,
\begin{equation}\label{2.12}
\delta H[\zeta]=\int_{\Sigma_t}(A^{ij}\delta g_{ij}+B^{ij}\delta \pi_{ij}),
\end{equation}
(without any surface terms) according to the ideology of Regge and Teitelboim~\cite{RT}.  Both the very existence
of $J[\zeta]$ (so, the very existence of $H[\zeta]$) and the form of it depend on the
boundary conditions.
 When
the constraint equations are fulfilled, $H[\zeta]$ reduce to $J[\zeta]$.  For the BTZ black hole solution the
 value of $H$  coincides with the black hole mass $M$.

Let us  define generators $L_n$ and $\bar{L}_n$ by relations $L_n=H[\zeta_n]$, $\bar{L}_n=H[\bar{\zeta}_n]$.
The Poisson brackets of these
generators  were calculated by Brown and Henneaux~\cite{BrHe}:
\begin{equation}\label{2.9}
[\hat{L}_n,\hat{L}_m]=(n-m)\hat{L}_{n+m}+{\frac{c}{12}}(n^3-n)\delta_{n+m,0},
\end{equation}
\begin{equation}\label{2.10}
[\hat{\bar{L}}_n,\hat{\bar{L}}_m]=(n-m)\hat{\bar{L}}_{n+m}+ {\frac{c}{12}}(n^3-n)\delta_{n+m,0},
\end{equation}
$[\hat{L}_n,\hat{\bar{L}}_m]=0$, where the constant $c$ is the central charge given by
\begin{equation}\label{2.4}
c=\frac{3\ell}{2G}.
\end{equation}
For the BTZ black hole Bekenstein-Hawking entropy is
\begin{equation}\label{2.3}
S_{BH}=\frac{2\pi r_+}{4G}=2\pi\sqrt{\frac{\ell^2 M}{2G}},
\end{equation}
where  $M=r_+^2/(8\ell^2G)$. This formula looks like the formula (\ref{1.7}) for the CFT entropy given in previous
subsection, where extension $\ell$ in Eq.~(\ref{2.3}) is analogous to parameter $b$ in (\ref{1.7}).

There is  the following relation between the energy and the Virasoro generators
\begin{equation}\label{2.14}
H=\frac 1\ell(H[\zeta_0]+H[\bar{\zeta}_0])= \frac 1\ell(L_0+\bar{L}_0).
\end{equation}
Let up to constraints $L_0=h$, $\bar{L}_0=\bar{h}$. If the energy of the system coincides
with the mass of the black hole, then, due to $h\leftrightarrow\bar{h}$ symmetry, $h=\bar{h}=M\ell/2$.

The main suggestion by Strominger is the existence of a quantum theory of gravity on $AdS_3$ and the coincidence of its
results with the results of classical theory given above.

In the semi-classical regime the spectrum of a
black hole can be treated as continuous. The classical black hole is a highly degenerate state $h+\bar h\gg c\gg 1$.
The degeneracy
$D$, $\bar{D}$ of operators $\hat{L}_n$, $\hat{\bar{L}}_n$ , can be found by using the Cardy formula~\cite{Cardy}
\begin{equation}\label{2.20}
\ln D\simeq 2\pi \sqrt{\frac{ch}{6}}=2\pi \sqrt{\frac{cL_0}{6}}.
\end{equation}
 The total degeneracy is
\begin{equation}\label{2.21}
\ln D+\ln\bar{D}=2\ln D\simeq 4\pi \sqrt{\frac{ch}{6}}= 2\pi\sqrt{\frac{\ell^2 M}{ 2G}} ,
\end{equation}
which is exactly the Bekenstein-Hawking entropy  of the BTZ black hole with mass $M$.

An obvious drawback of this derivation is certainly that it is limited to the $2+1$-gravity case. Though the idea
of transformation of  gauge degrees of freedom living on the boundary into dynamical variables due to boundary
conditions is attractive, but in the $2+1$-dimensional space-time it is irrelevant what part of the boundary
(outer or inner) is used while in $3+1$ case this is already not so. Therefore  boundary conditions should be
specified on the black hole horizon and not on spatial infinity.

\subsection{{Carlip I}}
\addcontentsline{toc}{subsection}{Carlip I}
Carlip~\cite{Carl:99a} invented a new method for statistical calculation of the black hole entropy in order to apply
the ideas by Strominger~\cite{Stro:97} discussed in the previous subsection to a general $n$-dimensional space-time,
without the restriction to $2+1$-dimensional case.

The metric of $n$-dimensional space-time may be taken as follows
\[
ds^2 = -N^2 dt^2 + f^2 (dr + N^r dt)^2 + \sigma_{\alpha\beta} (dx^\alpha + N^\alpha dt)(dx^\beta + N^\beta dt),
\]
where $ \alpha, \beta\ldots $ correspond to coordinates on a sphere $r={\rm const},  t={\rm const}$ and run over
values $ 1,2, \ldots, n-2 $. The function $N$ tends to zero on the  horizon of a black hole $r=r_+$ in such a way
that
\begin{equation}
N^2 = h (x^\alpha) (r-r_+) + O(r-r_+)^2.
\end{equation}

Further, the Hamiltonian looks like a linear combination of constraints $ \{ {\cal H}_t, {\cal H}_a \} $  plus,
probably, some surface integrals
\begin{equation}
H_0 [{\hat\xi}] = \int_\Sigma d ^ {n-1} x \, {\hat\xi} ^ \mu {\cal H} _ \mu,\qquad L[\hat\xi]=H_0[\hat\xi] + \oint
_ {\partial\Sigma} \ldots \ ,
\end{equation}
where the parameters of deformations of a constant time hypersurface (Lagrange multipliers for constraints) are
components of the decomposition for an infinitesimal space-time diffeomorphism  $ \xi^\mu $ over the  basis $\{
(1/N, -N^a/N), \partial/\partial x^a \}$:
\begin{equation}
{\hat\xi}^t = N\xi^t, \qquad {\hat\xi} ^a = \xi^a + N^a\xi^t.
\end{equation}
If one calculate a variation over $g_{ab} $ and $\pi^{ab}$ of that part of a Hamiltonian containing only
constraints (without the boundary terms), and  consider $\xi^\mu$ as not depending of canonical variables, one get
\begin{equation}
\begin{split}
\delta H_0&=\delta\int_\Sigma d^{n-1}x  {\hat\xi}^\mu {\cal H}_\mu =\\
&= \int_\Sigma d^{n-1}x\left(\frac{\delta H}{\delta g_{ab}} \delta g_{ab} +
\frac{\delta H}{\delta\pi^{ab}} \delta\pi^{ab} \right) -\\
&-{1\over 16\pi G} \oint_{\partial\Sigma} d^{n-2}x  \left\{ \sqrt{\sigma} \left(\sigma^{ac} n^b -\sigma^{ab}
n^c\right)\left({\hat\xi}^t\nabla_c\delta g_{ab}- \nabla_c {\hat\xi}^t\delta g_{ab} \right)+ 2
{\hat\xi}^a\delta\pi_a^{\ r} - {\hat\xi}^r \pi^{ab} \delta g_{ab}\right\},\\
\end{split}
\label{eq:variation}
\end{equation}
where $n^a$ is  a unit normal to the boundary at $t={\rm const}$, $K_{ab}$ is an extrinsic curvature tensor of a
constant time hypersurface, $\pi^{ab}=-f\sqrt{\sigma}(K^{ab}-g^{ab}K)$ are the momenta conjugate to the spatial
metric $g_{ab}$. The surface integral is taken over a boundary which should include both the horizon of a black
hole, and the spatial infinity. No boundary conditions are applied here. Up to  notations the formula coincides
with the similar formula from  Regge and Teitelboim work~\cite{RT}.

Then it is necessary to put boundary conditions for the canonical variables and for the parameters of
deformations. They are
\begin {enumerate}
\item to set in a phase space a domain ``close''  to a black hole solution;
\item to provide preservation of this domain under evolution determined by the
parameters of deformations.
\end {enumerate}
Carlip~\cite{Carl:99a} starts with the following boundary conditions at the horizon:
\[
f = {\beta h\over4\pi} N^{-1} + O(1), \qquad N^r = O(N^2), \qquad \sigma _{\alpha\beta} = O (1), \qquad N^\alpha =
O (1), \]
\[
(\partial_t -N^r\partial_r) g _ {\mu\nu} = O (N) g_{\mu\nu}, \qquad \nabla_\alpha N_\beta + \nabla_\beta N_\alpha
= O (N), \qquad
\partial_r N = O (1/N),
\]
\[
K_ {rr} = O (1/N^3), \qquad K _ {\alpha r} = O (1/N), \qquad K _ {\alpha\beta} =O (1),
\]
\[
{\hat\xi} ^r = O (N^2), \qquad {\hat\xi} ^t = O (N), \qquad {\hat\xi} ^ \alpha= O (1).
\]
What about a part of the
boundary arranged at spatial infinity we get rid of the corresponding contribution to the Hamiltonian variation by
considering parameters $\hat\xi $ as rapidly decreasing when $r\rightarrow\infty$.

Despite of the accepted boundary conditions the variation of Hamiltonian $H_0[\hat\xi]$ still contains some surface
contribution and to get rid of it is suggested~\cite{Carl:99a} to add  a surface integral of the following form
\begin{equation}
J [{\hat\xi}] = {1\over8\pi G} \oint _ {r=r _ +}  d ^ {n-2} x \Bigl\{n^a\nabla_a {\hat\xi} ^t\sqrt {\sigma} +
{\hat\xi} ^a\pi_a {} ^r + \left[n_a {\hat\xi} ^a K\sqrt {\sigma} \right] \Bigr\}. \label{a2}
\end{equation}
Later  the extra restrictions $K _ {rr} = 0 =K _ {\alpha\beta} $ are imposed on the exterior curvature
tensor.  As a result the last term in the variation formula (\ref{eq:variation}) becomes exactly zero and, in
fact,  the term written in square brackets of Eq.~(\ref{a2})  is not required by the Regge-Teitelboim ideology, just
opposite, it spoils the ``differentiability'' condition (in \cite{Carl:99a} this is masked by an unjustified
restriction $\delta K_{rr}/K_{rr}=O(N)$). So,
the variation of the improved Hamiltonian $L[\hat\xi]=H_0[{\hat\xi}] + J [{\hat\xi}]$ becomes
\begin{equation}
\delta L[\hat\xi] = \hbox{\em bulk terms} + {1\over8\pi G} \oint_{r=r_+}  d^{n-2} x \ \ \delta n^r\partial_r
{\hat\xi}^t  \sqrt{\sigma}.
\end{equation}

Then it is argued~\cite{Carl:99a} that,  first, we can calculate a Poisson bracket for
generators of two various deformations according to the following formula
\begin{equation} \left\{L [{\hat\xi}
_2], L [{\hat\xi} _1] \right\} = \delta _ {{\hat\xi} _2} L [{\hat\xi} _1],\label{eq:7}
\end{equation}
and second, that this equality remain valid after reduction, that is, after putting constraints to zero ``in strong sense''. It means, that the formula
(\ref{eq:7}) remains valid for surface integrals taken separately:
\begin{equation}
\hbox{\em boundary terms of} \left\{L[{\hat\xi}_2], L[{\hat\xi}_1] \right\} \approx  \hbox{\em boundary terms of}
\ \ \delta_{{\hat\xi}_2}H_0[\hat\xi_1]+\delta_{{\hat\xi}_2}J[{\hat\xi}_1]. \label{eq:8}
\end{equation}
In the same time the Poisson bracket (\ref{eq:7}) should have the following form
\begin{equation}
\delta_{{\hat\xi}_2}L[{\hat\xi}_1] = \left\{ L[{\hat\xi}_2], L[{\hat\xi}_1] \right\} = L\left[\{ {\hat\xi}_1,
{\hat\xi}_2 \}_{\hbox{\scriptsize SD}}\right] + K[{\hat\xi}_1,{\hat\xi}_2] \label{a13}
\end{equation}
where $K[{\hat\xi}_1,{\hat\xi}_2]$ is a possible central term in the algebra.  Here $\{ {\hat\xi}_1, {\hat\xi}_2
\}_{\hbox{\scriptsize SD}}$ is the Lie bracket for the algebra of surface deformations, given by
\begin{eqnarray}
\{ {\hat\xi}_1, {\hat\xi}_2 \}_{\hbox{\scriptsize SD}}^t &=& {\hat\xi}_1^a\partial_a{\hat\xi}_2^t -
{\hat\xi}_2^a\partial_a{\hat\xi}_1^t
\nonumber\\
\{ {\hat\xi}_1, {\hat\xi}_2 \}_{\hbox{\scriptsize SD}}^a &=& {\hat\xi}_1^b\partial_b{\hat\xi}_2^a -
{\hat\xi}_2^b\partial_b{\hat\xi}_1^a + g^{ab}\left( {\hat\xi}_1^t\partial_b{\hat\xi}_2^t -
{\hat\xi}_2^t\partial_b{\hat\xi}_1^t \right) . \label{a13a}
\end{eqnarray}
Eq.~(\ref{a13}) will be used  to compute the central charge.

As a result, the following expression was obtained for the surface contribution to the Poisson
bracket $\left\{L[{\hat\xi}_2],L[{\hat\xi}_1] \right\}$:
\begin{equation}
- {1\over8\pi G} \oint _ {r=r _ +}  d ^ {n-2} x \, \sqrt {\sigma} \left\{ { 1\over f^2}
\left[\partial_r(f{\hat\xi}_2^r)\partial_r{\hat\xi}_1^t-
\partial_r(f{\hat\xi}_1^r)\partial_r{\hat\xi}_2^t \right]
+ {1\over f} \partial_r\left[{\hat\xi} _1^r\partial_r {\hat\xi} _2^t - \delta _ {{\hat\xi} _2} {\hat\xi} _1^t
\right] \right\}. \label {a14}
\end{equation}
In further it is used  to realize the Virasoro algebra which central charge.

Then Carlip~\cite{Carl:99a} reduce the general formalism to the case of an axially symmetric black hole, with an adapted angular
coordinate $\phi$ such that $\partial_\phi g_{\mu\nu} = 0$ and consider a particular subalgebra of the surface deformation algebra with the following
properties:
\begin{enumerate}
\item The surface deformations are restricted to the {$r$--$t$} plane.
The $r$--$t$ plane has the central role in determining the entropy.
\item The diffeomorphism parameter $\xi^t = {\hat\xi}^t/N$ {``lives on the
horizon,''} in the sense that near $r=r_+$ it depends on $t$ and $r$ only in the combination $t-r_*$, where $fdr =
Ndr_*$ in the time-slicing such that $N_r=0$.  For the Kerr black hole, $t-r_*$ is essentially the standard
Eddington-Finkelstein retarded time, up to corrections of order $r-r_+$.
\item The lapse function $N$ is fixed at the horizon.  The horizon
is physically defined by the condition $N=0$, while our boundary term (\ref{a2}) is written at $r=r_+$; this
condition ensures that {the boundary remains at the horizon}.
\end{enumerate}
Condition $1$  imply that the diffeomorphism parameter $\xi^\phi$ has the form
\begin{equation}
\xi^\phi = -N^\phi\xi^t = -\frac{N^\phi}{N}{\hat\xi}^t . \label{b1}
\end{equation}
Condition $2$ requires that
\begin{equation}
\partial_r\xi^t = -\frac{f}{N}\partial_t\xi^t
\label{b2}
\end{equation}
at $r=r_+$, allowing us to write radial derivatives at the horizon in terms of time derivatives.  To impose
condition $3$, we can examine diffeomorphisms of $g^{tt} = -1/N^2$.  With initial coordinates chosen so that
$N_r=0$
\begin{equation}
\delta g^{tt} = 0 = \frac{2}{N^2}\left( \partial_t - N^\phi\partial_\phi\right)\xi^t + \frac{h}{N^4}\xi^r .
\label{b23}
\end{equation}
This structure suggests that the diffeomorphisms are separated into left-moving modes $\xi^t$ with
$\partial_t\xi^t = \Omega\partial_\phi\xi^t$, and right-moving modes $\tilde\xi^t$ with
$\partial_t{\tilde\xi}^t = -\Omega\partial_\phi{\tilde\xi}^t$, where $\Omega = -N^\phi(r_+)$ is the angular
velocity of the horizon.  Then
\begin{equation}
\xi^r = -\frac{4 N^2}{h}\partial_t\xi^t , \qquad {\tilde\xi}^r = 0. \label{b4}
\end{equation}
Now it is possible to write the left-moving modes at the horizon as
\begin{equation}
\xi^t_n = \frac{T}{4\pi}\exp\left\{ \frac{2\pi i n}{T} \left( t - r_* + \Omega^{-1}\phi \right) \right\} , \label{b5}
\end{equation}
where $T$ is an arbitrary period.   The normalization
 has been fixed by the requirement that the surface deformation algebra (\ref{a13a}) reproduce the
$\hbox{Diff}\,S^1$ algebra
\begin{equation}
\{ {\hat\xi}_m, {\hat\xi}_n \}_{\hbox{\scriptsize SD}}^t = i(n-m){\hat\xi}_{m+n}^t . \label{b6}
\end{equation}
Substituting the modes $\xi^t_n$ into the boundary term (\ref{a14}), we obtain
\begin{equation}
\delta_{{\hat\xi}_m} L[{\hat\xi}_n] = \hbox{\em bulk terms} +
  \frac{A}{8\pi G}\frac{\beta}{T}in^3\delta_{m+n} ,
\label{b8}
\end{equation}
where $A$ is the area of the boundary at $r=r_+$.  We can now use a trick of Brown and Henneaux to evaluate the
central term $K[{\hat\xi}_m, {\hat\xi}_n]$.  When evaluated on shell, the Hamiltonian and momentum constraints
vanish, so $H_0[{\hat\xi}]=0$.  Equation (\ref{a13}) thus reduces to a collection of boundary terms,
\begin{eqnarray}
\frac{A}{8\pi G}\frac{\beta}{T}in^3\delta_{m+n} = J[\{ {\hat\xi}_m, {\hat\xi}_n \}_{\hbox{\scriptsize SD}}] +
K[{\hat\xi}_m,{\hat\xi}_n] = \nonumber\\
=i(n-m)J[{\hat\xi}_{m+n}] + K[{\hat\xi}_m,{\hat\xi}_n] . \label{b9}
\end{eqnarray}
It is easily checked that
\begin{equation}
J[{\hat\xi}_p] = \frac{A}{16\pi G}\frac{T}{\beta}\delta_{p0} \label{b10}
\end{equation}
on shell.  Hence
\begin{equation}
K[{\hat\xi}_m,{\hat\xi}_n] =
 \frac{A}{8\pi G}\frac{\beta}{T}in(n^2-\frac{T^2}{\beta^2})\delta_{m+n} ,
\label{b11}
\end{equation}
the correct form for the central term of a Virasoro algebra with central charge
\begin{equation}
c = \frac{3A}{2\pi G}\frac{\beta}{T}. \label{b12}
\end{equation}

The results above imply that the quantum
states that characterize a black hole horizon must transform under a Virasoro algebra with central charge
(\ref{b12}).  This is sufficient to permit the use of  methods from CFT to count states. In particular, a CFT
with a central charge $c$ has a density of states $\rho(L_0)$ that grows
asymptotically as
\begin{equation}
\log\rho(L_0) \sim 2\pi\sqrt{ {c_{\hbox{\scriptsize eff}}L_0\over6} } , \label{c1}
\end{equation}
where $c_{\hbox{\scriptsize eff}}$ is an ``effective central charge''.  If the ground state is an eigenstate of
$L_0$ with eigenvalue zero---then $c_{\hbox{\scriptsize eff}} = c$.  Following Strominger, let us assume these
conditions are satisfied in quantum gravity.  Then from Eqs.~(\ref{b10}) and (\ref{b12}),
\begin{equation}
\log\rho(L_0) \sim {A\over4G} , \label{c2}
\end{equation}
giving the standard Bekenstein-Hawking entropy.  In general, right-moving modes may make an additional
contribution to the density of states, but  the central charge for those modes vanishes, so Eq.~(\ref{c2}) gives
the full entropy.

\bigskip
\noindent Carlip~\cite{Carl:99a} himself has provided some critical comments to his derivation:
\begin{itemize}
\item Like Strominger's derivation, this computation  uses symmetry arguments to derive the behavior of any
microscopic theory of black hole horizon states.  This is both good and bad: good because it provides a universal
explanation of black hole statistical mechanics, bad because it offers little further insight into quantum
gravity.
\item It also seems plausible that the description of black hole entropy here is related to the picture of
microscopic states as ``would-be pure gauge'' degrees of freedom that become dynamical at a boundary. The
existence of a central charge is an indication that the algebra of surface deformations has become anomalous, and
that invariance cannot be consistently imposed on all states.
\item The extremal black hole is typically characterized by a lapse function behaving as $N^2\sim(r-r_+)^2$ near the
horizon.  Such a configuration satisfies the boundary conditions assumed here, but in contrast to the non-extremal
result it gives $J[{\hat\xi}_0]=0$, i.e.  classically $L_0=0$.
\end{itemize}

\noindent{This approach has also been criticized for some other drawbacks by several authors:}~\cite{Solov,PaHo}

\begin{itemize}
\item The improved Hamiltonian remains non-differentiable.
\item Standard Poisson bracket calculation gives zero central charge.
 \item It is inapplicable to spherically symmetric case. The axial symmetry is required.
\end{itemize}

\subsection{Carlip II}
\addcontentsline{toc}{subsection}{Carlip II}
Being unsatisfied with the above mentioned contradictions in his derivation of the black hole entropy~\cite{Carl:99a}
Carlip proposed another approach~\cite{Carl:99b} to calculation of the surface terms in Poisson brackets.
It was based on a definition of Poisson brackets through the symplectic current and
covariant phase space technique.

First, let us consider a general diffeomorphism-invariant field theory in $n$ space-time dimensions with a Lagrangian ${\bf
L}[\phi]$, where $\bf L$ is viewed as an $n$-form and $\phi$ denotes an arbitrary collection of dynamical fields.
The variation of $\bf L$ takes the form \beq \delta{\bf L} = {\bf E}\cdot\delta\phi + d{\Th \ ,} \label{aa1} \eeq
where the field equations are given by ${\bf E}=0$ and the symplectic potential $\Th[\phi,\delta\phi]$ is an
$(n-1)$--form determined by the ``surface terms'' in the variation of $\bf L$. The symplectic current $(n-1)$--form
$\om$ is defined by \beq \om[\phi,\delta_1\phi,\delta_2\phi] =
   \delta_1\Th[\phi,\delta_2\phi] - \delta_2\Th[\phi,\delta_1\phi] ,
\label{aa2} \eeq and its integral over a Cauchy surface $C$, \beq \Omega[\phi,\delta_1\phi,\delta_2\phi] =
   \int_C \om[\phi,\delta_1\phi,\delta_2\phi]
\label{aa3} \eeq gives a presymplectic form on the space of solutions of the field equations.  This space, in
turn, can be identified with the usual phase space, and $\Omega$ becomes the standard presymplectic form of
Hamiltonian mechanics.

For any diffeomorphism generated by a smooth vector field $\xi^a$, one can define a conserved Noether current
$(n-1)$-form $\bf J$ by \beq {\bf J}[\xi] = \Th[\phi,{\cal L}_\xi\phi] - \xi\cdot{\bf L} \ , \label{aa4} \eeq where
${\cal L}_\xi$ denotes the Lie derivative in the direction $\xi$ and the dot $\cdot$ means contraction of a vector
with the first index of a form.  On shell, the Noether current is closed, and can be written in terms of an
$(n-2)$-form $\bf Q$, the Noether charge, as \beq {\bf J} = d{\bf Q} \ . \label{aa5} \eeq

Now consider a vector field $\xi^a$, and the corresponding generator of diffeomorphisms $H[\xi]$.  In the
covariant phase space formalism, Hamilton's equations of motion become \beq \delta H[\xi] =
\Omega[\phi,\delta\phi,{\cal L}_\xi\phi] . \label{aa6} \eeq It is easy to see that when $\phi$ satisfies the
equations of motion, \beq \om[\phi,\delta\phi,{\cal L}_\xi\phi]
   = \delta{\bf J}[\xi] - d(\xi\cdot\Th[\phi,\delta\phi]) ,
\label{aa7} \eeq so \beq H[\xi] = \int_{\partial C} ({\bf Q}[\xi] - \xi\cdot{\bf B} ) , \label{aa8} \eeq where the
$(n-1)$-form $\bf B$ is defined by the requirement that \beq \delta\int_{\partial C} \xi\cdot{\bf B}[\phi]
   = \int_{\partial C} \xi\cdot\Th[\phi,\delta\phi] .
\label{aa9} \eeq Given a choice of boundary conditions at $\partial C$, finding $\bf B$ is roughly equivalent to
finding the appro\-pri\-a\-te boundary terms for the Hamiltonian constraint in the standard ADM formalism of general
relativity.

The Lagrangian $n$-form of GR in $n$ space-time dimensions is \beq {\bf L}_{a_1\dots a_n} =
{1\over16\pi G}\epsilon_{a_1\dots a_n}R, \label{aa10} \eeq  yielding a symplectic potential $(n-1)$-form
 \beq
 \Th_{a_1\dots a_{n-1}}[g,\delta g] = {1\over16\pi G}
   \epsilon_{ba_1\dots a_{n-1}}\left( g^{bc}\nabla_c(g_{de}\delta g^{de})
   - \nabla_c\delta g^{bc}\right) .
\label{aa11} \eeq The corresponding Noether charge, evaluated when the vacuum field equations hold, is \beq {\bf
Q}_{a_1\dots a_{n-2}}[g,\xi] = -{1\over16\pi G}
   \epsilon_{bca_1\dots a_{n-2}}\nabla^b\xi^c .
\label{aa12} \eeq

In the absence of a boundary, the Poisson brackets of the generators $H[\xi]$ form the standard ``surface
deformation algebra'', equivalent on shell to the algebra of diffeomorphisms. On a manifold with boundary,
however, the addition of boundary terms can alter the Poisson brackets, leading to a central extension of the
surface deformation algebra.  That is, the Poisson algebra may take the form \beq \{H[\xi_1],H[\xi_2]\} =
H[\{\xi_1,\xi_2\}] + K[\xi_1,\xi_2], \label{bb1} \eeq where the central term $K[\xi_1,\xi_2]$ depends on the
dynamical fields only through their (fixed) boundary values.

Consider the Poisson brackets of the generators of diffeomorphisms in the covariant phase space formalism. Let
$\xi_1^a$ and $\xi_2^a$ be two vector fields, and suppose the fields $\phi$ solve the equations of motion (so, in
particular, the ``bulk'' constraints are all zero).  Denote by $\delta_{\xi}$ the variation corresponding to a
diffeomorphism generated by $\xi$. For the Noether current ${\bf J}[\xi_1]$, \beq
\delta_{\xi_2}{\bf J}[\xi_1] =
{\cal L}_{\xi_2}{\bf J}[\xi_1]
  = \xi_2\cdot d{\bf J}[\xi_1] + d(\xi_2\cdot{\bf J}[\xi_1])
  = d\left[ \xi_2\cdot(\Th[\phi,{\cal L}_{\xi_1}\phi]
  - \xi_1\cdot{\bf L}) \right] ,
\label{bb2} \eeq where the fact that $d{\bf J} = 0$ on shell was used.  Hence
\begin{eqnarray}
\delta_{\xi_2}H[\xi_1] &=& \int_C \delta_{\xi_2}{\bf J}[\xi_1]
   - d(\xi_1\cdot\Th[\phi,{\cal L}_{\xi_2}\phi]) \nonumber\\
&=& \int_{\partial C} \left(\xi_2\cdot\Th[\phi,{\cal L}_{\xi_1}\phi]
   - \xi_1\cdot\Th[\phi,{\cal L}_{\xi_2}\phi]
   - \xi_2\cdot(\xi_1\cdot{\bf L}) \right) .
\label{bb3}
\end{eqnarray}

Then Brown-Henneaux~\cite{BrHe} trick is used.  Since equation above is evaluated on
shell, the ``bulk'' part of the generator $H[\xi_1]$ on the l.h.s., which is a sum of
constraints, vanishes.  Hence  l.h.s.  can be interpreted as the variation $\delta_{\xi_2}J[\xi_1]$,
where $J$ is the boundary term in the constraint.   On the other hand, the Dirac bracket $\{J[\xi_1],J[\xi_2]\}^*$
means precisely the change in $J[\xi_1]$ under a surface deformation generated by $J[\xi_2]$; that is, \beq
\delta_{\xi_2}J[\xi_1] = \{J[\xi_1],J[\xi_2]\}^* . \label{bb4} \eeq Comparing Eq.~(\ref{bb1}), evaluated on
shell, we see that \beq K[\xi_1,\xi_2] = \delta_{\xi_2}J[\xi_1]
   - J[\{\xi_1,\xi_2\}] ,
\label{bb5} \eeq where $\delta_{\xi_2}J[\xi_1]$ is given by  (\ref{bb3}). Given a suitable set of boundary
conditions, this allows to determine the central term $K[\xi_1,\xi_2]$.

In particular, for GR without matter sources, the Lagrangian $\bf L$ vanishes on shell, and  r.h.s. of
 (\ref{bb3}) can be estimated from  (\ref{aa11}).  We get
 \begin{eqnarray} \{J[\xi_1],J[\xi_2]\}^* = {1\over16\pi
G}
   \int_{\partial C}\epsilon_{bca_1\dots a_{n-2}}\left[
   \xi_2^b\nabla_d(\nabla^d\xi_1^c - \nabla^c\xi_1^d) -\right.\nonumber\\
   -\left.   \xi_1^b\nabla_d(\nabla^d\xi_2^c - \nabla^c\xi_2^d) \right] .
\label{bb6} \end{eqnarray}

Consider an $n$-dimensional space-time $M$ with boundary $\partial M$, such that a neighborhood of $\partial M$
admits a Killing vector $\chi^a$ that satisfies $\chi^2 = g_{ab}\chi^a\chi^b = 0$ at $\partial M$.  In practice,
it is useful to work at a ``stretched horizon'' $\chi^2=\epsilon$, taking $\epsilon$ to zero at the end of the
computation.  Near this stretched horizon, one can define a vector orthogonal to the orbits of $\chi^a$ by \beq
\nabla_a\chi^2 = -2\kappa\rho_a , \label{cc1} \eeq where $\kappa$ is the surface gravity at the horizon.  Note
that \beq \chi^a\rho_a = -{1\over\kappa}\chi^a\chi^b\nabla_a\chi_b = 0 . \label{cc2} \eeq At the horizon, $\chi^a$
and $\rho^a$ become null, and the normalization in Eq.~(\ref{cc1}) has been chosen so that
$\rho^a\rightarrow\chi^a$. Away from the horizon, however, $\chi^a$ and $\rho^a$ define two orthogonal directions.

Then we focus on vector fields of the form
\beq
\xi^a = R\rho^a + T\chi^a .
\label{cc4}
\eeq
The corresponding diffeomorphisms are, in a reasonable sense,
deformations in the ``$r$--$t$ plane,'' which are known to play a
crucial role in the Euclidean approach to black hole thermodynamics.

Now, given any one-parameter group of diffeomorphisms satisfying
boundary conditions, it is possible to check that
\beq
\{ \xi_1,\xi_2  \}^a = (T_1DT_2 - T_2DT_1)\chi^a +
  {1\over\kappa}{\chi^2\over\rho^2}D(T_1DT_2 - T_2DT_1)\rho^a .
\label{cc8}
\eeq
This is isomorphic to the standard algebra of diffeomorphisms of the
circle or the real line.  The question  is whether the algebra
of constraints merely reproduces this $\hbox{\it Diff\,}S^1$ or
$\hbox{\it Diff\,}{\bf R}$ algebra, or whether it acquires a central
extension.
Finally, one finds that
\beq
\{J[\xi_1],J[\xi_2]\}^* = -{1\over16\pi G}\int_{\cal H}
   {\hat\epsilon}_{a_1\dots a_{n-2}}\left[
   {1\over\kappa}(T_1D^3T_2 - T_2D^3T_1) - 2\kappa(T_1DT_2-T_2DT_1)
   \right] ,
\label{cc13}
\eeq
where terms of order $\chi^2$ have been omitted.

This expression has the characteristic three-derivative structure of
the central term of a Virasoro algebra.  We must also compute the surface term $J[\{\xi_1,\xi_2\}]$ of
the Hamiltonian to obtain the complete expression for the central term
in the constraint algebra.  This Hamiltonian
consists of two terms.  The first is straightforward to compute, one finds that
\beq
{\bf Q}_{a_1\dots a_{n-2}} = {1\over16\pi G}
   {\hat\epsilon}_{a_1\dots a_{n-2}}\left(
   2\kappa T - {1\over\kappa}D^2T \right) + O(\chi^2) .
\label{cc14}
\eeq
The second term is more complicated, it is shown that it makes no further contribution.
Hence  we obtain a central
term
\beq
K[\xi_1,\xi_2] =  {1\over16\pi G}\int_{\cal H}
   {\hat\epsilon}_{a_1\dots a_{n-2}}{1\over\kappa} \left(
   DT_1D^2T_2 - DT_2D^2T_1 \right) ,
\label{cc15}
\eeq
and a centrally extended constraint algebra
\beq
\{ J[\xi_1], J[\xi_2] \}^* = J[\{\xi_1,\xi_2\}]
   + K[\xi_1,\xi_2] .
\label{cc16}
\eeq
A
one-parameter group of diffeomorphisms satisfying orthogonality that
closes under the brackets (\ref{cc8}) is then given by
\beq
T_n = {1\over\kappa}\exp\left\{ in\left( \kappa v + \sum_\alpha
   \ell_{\alpha}(\phi_{(\alpha)} - \Omega_{(\alpha)}v) \right)\right\} ,
\label{dd6} \eeq where the $\ell_{\alpha}$ are arbitrary integers, at least one of which must be nonzero, and the
normalization has been chosen so that \beq \{ T_m, T_n \} = -i(m-n)T_{m+n}
\ .\eeq Assuming the orthogonality
relation, it is easy to see that the algebra (\ref{cc16}) is now a conventional Virasoro algebra.  The microscopic
degrees of freedom, whatever their detailed characteristics, must transform under a representation of this
algebra.  But as Strominger observed~\cite{Stro:97}, this means that these degrees of freedom have a conformal
field theoretic description, and powerful methods from conformal field theory are available to analyze their
properties.

In particular, we can now use the Cardy formula to count states.
If we consider modes of the form (\ref{dd6}), the central term (\ref{cc15})
is easily evaluated:
\beq
K[T_m,T_n] = -{iA\over8\pi G}m^3\delta_{m+n,0} ,
\label{dd8}
\eeq
where $A$ is the area of the cross section $\cal H$.  The algebra
(\ref{cc16}) thus becomes
\beq
i\{J[T_m],J[T_n]\}^*
  = (m-n)J[T_{m+n}] + {A\over8\pi G}m^3\delta_{m+n,0} ,
\label{dd9} \eeq which is the standard form for a Virasoro algebra with central charge \beq {c\over12} =
{A\over8\pi G} . \label{dd10} \eeq The Cardy formula also requires that we know the value of the boundary term
$J[T_0]$ of the Hamiltonian.  This can be computed from Eq.~(\ref{cc14}): \beq J[T_0] = {A\over8\pi G}.
\label{dd11} \eeq

The Cardy formula then tells us that for any conformal field theory that provides a representation of the Virasoro
algebra (\ref{dd9})---modulo certain assumptions ---the number of states with a given eigenvalue $\Delta$ of
$J[T_0]$ grows asymptotically for large $\Delta$ as \beq \rho(\Delta) \sim \exp\left\{ 2\pi
\sqrt{{c\over6}\left(\Delta-{c\over24}\right)} \right\} . \label{dd12} \eeq Inserting Eqs.~(\ref{dd10}) and
(\ref{dd11}), we find that \beq \log\rho \sim {A\over4G} \ , \label{dd13} \eeq giving the expected behavior of the
entropy of a black hole.

Despite the strong desire by Carlip to avoid technical errors this approach has also been criticized~\cite{Guo}.
Another calculation~\cite{Guo} gives zero result for the central charge. The difference is in the choice of basis
and it is observed that the basis taken by Carlip is singular on the horizon.

\subsection{Carlip III}
\addcontentsline{toc}{subsection}{Carlip III}

A new method  was proposed and applied to  2-dimensional dilaton gravity~\cite{Carl:02}. The main idea seems to be that
the black hole horizon
should contribute directly to the symplectic form and so to the Poisson brackets.
\bigskip

For the discussion of details and criticism we address reader to paper~\cite{KKP:04}. It seems to be appropriate to
remind here that
a surface contribution to the symplectic form and Poisson brackets
have been observed and studied in \cite{Soloviev:92,Soloviev:93}.

\section{Other approaches to statistical derivation of black hole entropy}
\addcontentsline{toc}{section}{Other approaches to statistical derivation of black hole entropy}
\subsection{String theory}
\addcontentsline{toc}{subsection}{String theory} String theory becomes related to black holes due to supergravity
which is its low energy limit. The special interest is attracted to black hole solutions which do not get
corrections under quantization. These are so-called BPS solitons. String theory methods alow to vary the coupling
constant keeping entropy unchanged. For small string coupling constant a black hole behaves as a point particle in
flat space-time.

We consider as the first work in this direction the paper by Strominger and Vafa~\cite{StVa:96}, where Bekenstein-Hawking
formula has been derived for one class of 5-dimensional extremal black holes. This result has been reached by a calculation
 of degeneration degree for BPS soliton states. The  degeneration degree was estimated  by topological methods and in
 leading order the statistical theory gave the following formula
\beq S_{stat}=2\pi \sqrt{Q_H({1\over 2}Q_F^2+1)},\label{StVa_stat} \eeq where $Q_H$ is the axion charge and $Q_F$
--- the electrical charge. In the same time proceeding from low energy effective action one could derive the
Bekenstein-Hawking entropy as \beq S_{BH}=2\pi \sqrt{Q_HQ_F^2 \over 2}.\label{StVa_BH} \eeq The obtained results
(\ref{StVa_stat}) and (\ref{StVa_BH}) coincide when charges are large. $Q_H$ and $Q_F^2/2$ are here integers.

The essential content of this pioneer work was the exploration of five noncompact dimensions and $N=4$
supersymmetry. 5-dimensional space occurred the first space giving a non-trivial result for the entropy. To make
the horizon area different from zero it was necessary to have nonzero values for both charges  $Q_H$ and $Q_F$.

Of course, there were a lot of following publications that considerably strengthened first results. So, Horowitz
and Strominger in article~\cite{HoSt:96} generalized calculations of~\cite{StVa:96} onto near extremal black holes
(in the first order deviation from extremal ones). The excitations of extremal black holes were identified with
the string states. This allowed to study also a process of black hole evaporation in the framework of perturbation
theory. It was found that unitarity is preserved in this process.  5-dimensional black hole was replaced  by
6-dimensional black string wrapped around a compactified dimension (1-circle). Then the black hole states were
described as degrees of freedom inhabiting this circle.

The extremal black hole has 2 sorts of charge: one charge is responsible for
the number of degrees of freedom and another --- for their right momenta. Left momenta are zero in case of the extremal black hole. It is supposed that
a non-extremal black hole would have also nonzero left momenta. The extremal solution with zero right momenta has zero horizon area, i.e it is non-degenerate. But if there are nonzero right momenta, then the black string though staying extremal gets a nonzero horizon area.

The dimensional reduction transforms 6-dimensional black string into 5-dimensional black hole such as described in
paper~\cite{StVa:96}. If one adds left momenta then the black string becomes not extremal and compactifies to a
non-extremal black hole. Calculating the number of black string states Horowitz and Strominger has obtained a
precise correspondence with the black string entropy formula.

The following work by  Maldacena and Strominger~\cite{MaSt:96} has allowed to go from 5-dimensional extremal black
holes to 4-dimensional extremal ones. The methods of works discussed above were based on calculations with
D-branes. But 4-dimensional black hole with a nonzero horizon area could not be constructed of D-branes only. It
was necessary to introduce a new object called a symmetric 5-brane or Kaluza-Klein monopole and to apply a new
technique of calculations. This was done in paper~\cite{MaSt:96}.

Exploiting the standard formula for $(1+1)$-entropy
\beq
S=\sqrt{\pi (2N_B+N_F)EL\over 6},\label{MaSt_std}
\eeq
where $N_B~~(N_F)$ is a number of boson (fermion) species moving in
right direction,  $E$ is a total energy, and $L$ is characteristic length,
and also using relations $N_B=N_F=4Q_2Q_6$, $E=2\pi
n/L$, one might get for thermodynamical limit of large  $n$ to an independent of $L$ result:
\beq
S_{stat}=2\pi\sqrt{Q_2Q_6n} \ .
\eeq
The Bekenstein-Hawking entropy is as follows
\beq S_{BH}=2\pi\sqrt{Q_2Q_6nm} \ , \eeq
where $m$ is the axion charge of a symmetric 5-brane wrapped around  $Y\times S^1.$ As this charge is absent in the
compactification  $\hat S^1$, so  $S_{BH}$ is always proportional to  $(charge)^2$ when estimated from
the leading order low energy effective action. This is quite different from
5-dimensional case when it is proportional to $(charge)^{3/2}$.
As  $S_{stat}$ is proportional to $(charge)^{3/2}$, it appears in leading order only in 5 dimensions and stays an
invisible correction of higher order
in 4 dimensions.

In order to get a nonzero area in 4 dimensions it is required to add $m$ 5-branes, wrapping
 $Y\times S^1$. These $m$
5-branes may be placed anywhere on $\hat S^1$. Each 2-brane has to intersect
 all $m$ 5-branes along  $S^1$. A 2-brane can be broken and its ends separate in $Y$ when crossing a 5-brane.  Hence the
$Q_2$ toroidal 2-branes break up into $mQ_2$ cylindrical 2-branes, each of which is bounded by a pair of 5-branes.
The momentum-carrying open strings now carry an extra label describing which pair of 5-branes they lie in between.
The number of species becomes $N_B=N_F=4mQ_2Q_6$. Inserting this into (\ref{MaSt_std}) together with $E=2\pi n/L$
we obtain \beq S_{stat}=2\pi\sqrt{Q_2Q_6nm},\eeq In agreement with the semiclassical result  for $S_{BH}$.

For the $N=4$ case
 there are, in general,  28 electric charges $\vec Q$ and 28 magnetic
charges $\vec P$. In this notation $2Q_2Q_6=\vec P^2$ and $ 2 nm=\vec Q^2$. Duality implies that the entropy
depends only on $\vec P^2, ~\vec Q^2$ and $\vec Q \cdot \vec P$. The general formula for the Bekenstein-Hawking
entropy is \beq S_{BH}=\pi \sqrt{ \vec P^2 \vec Q^2-(\vec Q \cdot \vec P)^2}. \eeq For the discussed case the last
term vanishes.

{One may ask the following questions:}
\begin{itemize}
\item What is the nature of these degrees of freedom responsible for black hole entropy?

\item Where they are located?

\item Is it possible to treat non-extremal black holes by this method?
\end{itemize}
They are still open and should be answered if the string theory pretends to be
a theory of everything.

\subsection{Loop quantum gravity}
\addcontentsline{toc}{subsection}{Loop quantum gravity} In the Ashtekar approach~\cite{ABCK:97} first a phase space is
constructed which describes a space-time con\-tain\-ing an isolated non-rotating black hole. Then it is canonically
quantized. As a result  a black hole sector in the non-perturbative  canonical quantum gravity is obtained. Quantum
states determining the black hole horizon geometry are separated. Just they are responsible  for the
Bekenstein-Hawking entropy. It is found that these black hole degrees of freedom can be described by Chern-Simons
field theory on the horizon. It is shown that the statistical mechanical entropy is proportional to
the horizon area for large non-rotating black holes. The constant of proportionality depends upon the Immirzi
parameter~\cite{ImmPar}, which fixes the spectrum of the area operator in loop quantum gravity; an appropriate choice of
this
parameter gives the Bekenstein-Hawking formula $S = A/4\ell_{Pl}^2$. The same choice of the Immirzi parameter also gives
the
similar result for black holes carrying electric or dilatonic charge. An important advantage of this method is that black
holes considered are not necessary  extremal  or near extremal, and also they are certainly 4-dimensional (not 5-dimensional\ !).

Non-perturbative techniques of the Ashtekar approach have led to a quantum theory of geometry in which operators
corresponding to lengths,
areas and volumes have discrete spectra~\cite{area}.  Of particular interest are the spin network states associated with
graphs in 3-space with edges labelled by spins $j = {1\over 2},1, \dots$ and vertices labelled by intertwining
operators. If a single edge punctures a 2-surface transversely, it contributes an area proportional to
$\sqrt{j(j+1)}$.

A sector of the classical phase space corresponding to an
isolated, uncharged, non-rotating black hole can be described starting from  a manifold with a boundary. This space-time
manifold has the outer boundary  ${\cal I}$ and the inner boundary as ${\cal H}$.  ${\cal I}$ corresponds to the
infinitely far region and  ${\cal H}$~--- to the black hole horizon. Hamiltonian variables are a soldering form
$\sigma_a^{AA'}$ for $\SL(2,\C)$ spinors and an $\SL(2,\C)$ connection $A_{aA}{}^{B}$.
Metric  $g_{ab} = \sigma_a^{AA'} \sigma_{bAA'}$ is the Lorentzian space-time metric and $A_{aA}{}^B$
is the self-dual connection operating on unprimed spinors only.
On ${\cal I}$ fields are required to satisfy
the standard asymptotically flat boundary conditions. The conditions on $\H$, on the other hand, are specially elaborated
by Ashtekar's group.
The key requirements are: i) ${\cal H}$ should be a null surface with respect to the metric $g_{ab}$; ii) On a `finite
patch' $\Delta$ of $\H$, the area of any cross-section should  be a constant, $A_{\B}$, the Weyl spinor be of Petrov type
2-2 and its only non-zero component, $\Psi_2$,  should be given by $\Psi_2 = {2\pi}/A_{S}$; and, iii) the 2-flats on
$\Delta$, orthogonal to the two principal null directions of the Weyl tensor span 2-spheres and the pull-back of
the connection $A_a$ to these 2-spheres  should be real.

 In short, condition ii) implies that there is no gravitational radiation falling into $\Delta$ (`isolated'
 black hole) while the first part of iii) implies that it is `non-rotating'.
These boundary conditions have been taken from the Schwarzschild black hole horizon. But here space-time is  not
required to be static and gravitational waves are admitted far from the black hole. Therefore, the phase space is
infinite dimensional. Nevertheless, these boundary conditions are sufficient for the existence of a surface
integral providing differentiability of the action, so the Regge--Teitelboim criterion~\cite{RT} is fulfilled. The
required surface term is just the Chern-Simons action:
\begin{equation} \label{action}
S(\sigma,A) = -\,{i\over 8\pi G}\int_\M\,{\rm Tr}\,
\left(\Sigma\wedge F\right)
- {i\over 8\pi G}{A_{\B}\over 4\pi}\int_{\Delta}{\rm Tr}
\left(A\wedge dA + {2\over 3}A\wedge A\wedge A\right),
\end{equation}
where $\Sigma_{ab}^{AB} = 2 \sigma_{[a}{}^{AA'}\sigma_{b]A'}{}^B$, and
$F_{abA}{}^B$ is the curvature of the connection $A$.
In the Hamiltonian formalism following from (\ref{action})  canonical variables are the restrictions of
$\Sigma$ and $A$ to the spatial hypersurface $M$ with a boundary
$\B$. Here a difficulty arises due to the fact that the restriction to
$M$ of the self-dual connection $A$ is a complex valued $\SU(2)$
connection. Then it is better to make a transition to
real variables: $A_a = \Gamma_a - i K_a$, where $\Gamma$ is the
3-dimensional spin connection compatible with the triad field and $K$
is the extrinsic curvature of $M$. Here the Immirzi~\cite{ImmPar} parameter $\i$  appears.
It is a positive real number and real phase space variables are
$\A_a := \Gamma_a - \i K_a$ and $\E_{ab} := (1/\i) \Sigma_{ab}$.
 Then the boundary conditions imply
that the phase space consists of real fields $(\A_a, \E_{ab})$ that
are asymptotically flat at infinity and satisfy the following
condition at $\B$:
\be \label{bc}
{}^\gamma\!\underline{F}_{ab}^{AB} = - \frac{2\pi\gamma}{A_{\B}}\,\,
{}^\gamma\!\underline{\Sigma}_{ab}^{AB} \, ,
\ee
where underlining means pull-backs to $\B$. The restriction of $\A_a$ to $\B$ yields a reducible connection
satisfying $D_a r = 0$ for some radial internal vector
$r$. It is suitable to fix $r$ on $\B$ using the $\SU(2)$ gauge freedom. Then the
gauge group on the boundary is reduced to $\U(1)$ and only the $r$
component of~(\ref{bc}) is non-trivial.

On the real phase space, the symplectic structure
corresponding to (\ref{action}),  in addition to the familiar volume term, contains a surface contribution
(see \cite{Soloviev:92} for the first evidence of a surface term in Ashtekar's symplectic form) which coincides with the
symplectic structure of the Chern--Simons theory:
\begin{equation}\label{ss}
\Omega
 =
{1\over 8\pi G} \int_M {\rm Tr}
\left [ \delta\E\wedge\delta\A' - \delta\E'\wedge\delta\A \right ]
- \frac{A_S}{16\pi^2\i G} \oint_\B  \,{\rm Tr}
\left [ \delta\A\wedge\delta\A' \right ].
\end{equation}
There are bulk degrees of freedom,
corresponding to gravitational waves far away from $\Delta$, which
should not be taken into account as genuine black hole degrees of
freedom.  The boundary degrees of freedom describing the geometry of
the horizon $\B$ are
degrees of freedom `living on the horizon' and responsible  for the
entropy.

In the classical theory  the bulk
and boundary degrees of freedom cannot be separated: all fields on $\B$
are determined by fields in the interior of $M$ as they are required to be continuous.
However, in the quantum theory, the fields describing geometry become
discontinuous, and the fields on
$\B$ are no longer determined by fields in $M$; in this case there are
independent degrees of freedom `living' on the boundary.  These
surface degrees of freedom are the ones that account for black hole
entropy in this approach.

To quantize the theory a Hilbert space $\H^V$ of
`volume' states and a Hilbert space $\H^S$ of `surface' states are constructed, and
then constraints are imposed on $\H^V \otimes \H^S$ to obtain the space of
physical states.  $\H^V$ is taken to consist of
square-integrable functions on the space of generalized $\SU(2)$
connections  on $M$ modulo gauge transformations that are
 identity on $\B$.  The form of the Hilbert space $\H^S$ of surface
states is motivated by the fact that in the quantum theory it is required to impose the boundary condition
(\ref{bc}) as an operator equation. That is, given a spin network state $\Psi_V$ in $\H^V$ and a state $\Psi_S$ in
$\H^S$, the quantum version of the $r$ component of equation (\ref{bc}) should read \be\label{qbc}
(1{\textstyle\bigotimes} {2\pi\gamma\over A_{\B}}\hat{ \underline{F}}_{ab}\cdot r + \hat{\underline{\Sigma}}_{ab}
\cdot r {\textstyle\bigotimes} 1)\, \Psi_V {\textstyle\bigotimes} \Psi_\B\, = 0. \ee The structure of this
equation implies that $\Psi_V$ and $\Psi_\B$ should be eigenstates of $\hat{\underline{\Sigma}}_{ab}\cdot r$ and
$\hat{\underline{F}}_{ab}\cdot r$ respectively. The `polymer nature' of quantum geometry in $M$ implies that
eigenvalues of ${\hat{\underline\Sigma}_{ab}\cdot r}$ are distributional, given by \be 8\pi \l^2\sum_{i}\, j_i
\delta^2(x,p_i) \eta_{ab}\, \l^2 \ee
for some points $p_i$ on $\B$, where $j_i$ are half-integers, $\delta^2$ is
the delta distribution on $\B$, $\eta_{ab}$ the Levi-Civita density on $\B$ and $\l$ the Planck length. Therefore,
(\ref{qbc}) implies that the surface states $\Psi_S$ have support only on generalized connections that are
everywhere flat except at a finite number of points $p_i$. It turns out that such generalized connections can be
identified with ordinary connections with distributional curvature.  Since the surface symplectic structure is
that of Chern--Simons theory, for any fixed choice
$$\P=\{(p_1,j_{p_1}),\ldots,(p_n,j_{p_n})\}$$
of points in $\B$ labelled by spins,  $\H^S$ should have a subspace given by the space of states of $\U(1)$
Chern--Simons theory on a sphere with punctures $p$ labelled by spins $j_p$. The total space $\H^S$ is the direct
sum of these subspaces.

Note now that $(k/2\pi) \hat{\underline F}$ is the generator of internal rotations in Chern--Simons theory. Thus,
the meaning of (\ref{qbc}) turns out to be rather simple: it ensures that the volume and surface states are
`coupled' in precisely the correct way so that the total state is invariant under $\U(1)$ internal rotations at
$\B$. The remaining constraints require that the states be invariant under diffeomorphisms of $M$ that leave $\B$
invariant and under motions generated by the Hamiltonian constraint smeared with any lapse field that vanishes at
$\B$. Thus for each set $\P$ of finitely many punctures $p$ labelled by spins $j_{p}$, there is a subspace
$H^V_\P$ of volume states having a basis given by open spin networks whose edges intersect $\B$ only at these
punctures, where they are labelled by the spins $j_{p}$. Similarly there is a subspace $H^\B_\P$ consisting of
quantum states of $\U(1)$ Chern-Simons theory on the punctured surface $\B$.  The total physical Hilbert space is
given by:
$$\H_{\rm phy} =
{\bigoplus_\P\, \left [\H^V_\P \otimes \H^S_\P \right ] \over{\rm Gauge}}
\ ,$$
where `Gauge' means internal $\SU(2)$
rotations that reduce to $\U(1)$ on $\B$, diffeomorphisms preserving $\B$, and the motions generated by the
Hamiltonian constraint. To proceed further it is
necessary to make an assumption about the quantum dynamics: let there is at least one solution of this constraint
in $\H^V_P \otimes \H^S_\P$ for any set $\P$ of punctures labelled by spins.

It is required to consider only states of the horizon of a black hole with
area $A$.  Thus  the `volume' states are traced over to construct a
density matrix $\rho_{\rm bh}$ describing a maximal-entropy mixture of
surface states for which the area of the horizon lies in the range $A_{\B}
\pm \l^2$.  The statistical mechanical black hole entropy is then
given by $S_{\rm bh} = - {\rm Tr} \rho_{\rm bh} \ln \rho_{\rm bh}$.
This can be computed by counting states: $S_{\rm bh}
= \ln N_{\rm bh}$ where $N_{\rm bh}$ is the number of Chern-Simons
surface states satisfying the area constraint.

The eigenvalues of the area operator  are given by the following formula~\cite{area}: \be \label{ev} 8\pi\i \l^2
\sum_p \sqrt{j_p(j_p+1)} \ , \ee
where $j_p$ are the spins labelling the punctures. Using this and the result from
Chern-Simons theory that for a large number of punctures the dimension of $\H^S_\P$ grows as
\begin{equation}
{\rm dim} \H^S_\P \,\, \sim \,\, \prod_{j_p\in\P} (2j_p+1) \ ,
\end{equation}
it is possible to calculate the entropy.  For large $A$ it is
given by
\begin{eqnarray}\label{ent}
S_{\rm bh} = \frac{\i_0}{4\l^2\i}\,  A_{\B} ,\quad
\i_0 = \frac{\ln{2}}{\pi\sqrt{3}}.
\nonumber
\end{eqnarray}
Thus, in the
limit of large area, the entropy is proportional to the area of the
horizon.  If  $\i$ is put equal to $\i_0$, then the calculation based on statistical mechanics gives a result
coinciding with  the Bekenstein-Hawking formula.

As it is stressed by authors~\cite{ABCK:97}, the advantages of their method are 1) that black holes considered are
not necessary  extremal  or near extremal, and 2) that dimension of space-time has just the observable value.
Nevertheless, as one can see from above,  Ashtekar's approach to quantum gravity faces with some principal
difficulties. First, they seem to be coming from the complex structure of the phase space, second, from the fact
that  space-time dimension here can not be easily changed in Ashtekar's formalism. Probably, other difficulties,
such as the absence of proof that  the Hamiltonian constraint  solutions really exist, and the presence of a free
parameter (Immirzi parameter $\i$),
 are consequences of these two.
 
 \newpage
 
\section{Conclusions}
\addcontentsline{toc}{section}{Conclusions}
\begin{itemize}
\item There is a strong belief that the problem of statistical derivation of the Bekenstein-Hawking formula for
black hole entropy {can be solved at the present level of our physical knowledge}.

\item It is very probable that the path to solution consists in exploiting {conformal invariance} near the black hole
horizon.

\item But now this problem is {far from being completely solved}.

\item Every successful approach to the problem could help us to understand {what quantum gravi\-ty~is}.
\end{itemize}
It would be unrealistic
to try to cover all approaches to statistical derivation of the black hole entropy here, as an example let us mention
rather radical ideas
 appeared quite recently in works by V.V. Kiselev~\cite{Kiselev}.
We would like to address interested readers to more detailed reviews of the subject~\cite{Fursaev,Giacomini,Padman}.

\end{document}